\documentstyle[aps,prb,twocolumn,epsfig]{revtex}

\newcommand{\bscco}{{Bi$_2$Sr$_2$Ca$_2$Cu$_2$O$_{8+\delta}$}}

\begin{document}
\twocolumn[\hsize\textwidth\columnwidth\hsize\csname
@twocolumnfalse\endcsname

\title{Conductivity Due to Classical Phase Fluctuations
in a Model For High-T$_c$ Superconductors}
\author{S.\ Barabash$^1$, D.\ Stroud$^1$, and I.-J.\ Hwang$^2$ }
\address{$^1$Department of Physics,
The Ohio State University, Columbus, Ohio 43210}
\address{$^2$ IBM Almaden Research Center, San Jose, CA 95120} 
\date{\today}

\maketitle

\begin{abstract}
We consider the real part of the conductivity,
$\sigma_{1;\alpha\alpha}(\omega)$,
arising from classical phase fluctuations in a model for
high-T$_c$ superconductors.  
We show that $\int_0^\infty\sigma_{1;\alpha\alpha}d\omega \neq 0$ 
below the superconducting transition temperature $T_c$, 
provided there is some quenched disorder in the system.  
Furthermore, for a fixed amount of quenched disorder, this integral
at low temperatures is proportional to the zero-temperature
superfluid density, in agreement with experiment.  We calculate 
$\sigma_{1;\alpha\alpha}(\omega)$ explicitly for a model of overdamped
phase fluctuations. 
\end{abstract}

\draft 
\vskip1.5pc] 
 
\par
%\begin{twocolumn}
Because of their high transition temperatures, small coherence lengths, 
and low superfluid densities, the cuprate 
superconductors are strikingly influenced by phase fluctuations of the 
superconducting order parameter.   Such fluctuations are largely
responsible for flux lattice melting\cite{nelson} and vortex 
glass\cite{fisher} transitions in a finite magnetic field.   
In addition, they strongly affect the
zero-field transition\cite{salamon}, and possibly also the 
superconducting transition temperature itself in underdoped 
materials\cite{emery}.  
Phase fluctuations also influence the transport 
properties of the high-T$_c$ materials.  For example, 
the finite-frequency conductivity shows a fluctuation-induced
peak near $T_c$\cite{anlage}. 

Recent measurements in the most anisotropic high-T$_c$ materials, such
as \bscco, (BSCCO) have found that the
real part of the conductivity, $\sigma_1(\omega)$, not only has a peak
near $T_c$ but also remains quite large even
far below $T_c$\cite{corson}.  Published
measurements are available typically at frequencies in the
30-200 GHz range\cite{corson,lowfreq}. In some cases, $\sigma_1(\omega, T)$
at such low temperatures exceeds the peak values observed near
$T_c$\cite{orenstein}.
These large values occur over a broad concentration range varying
from overdoped to optimally doped to underdoped.  Finally, large
low-temperature values of $\sigma_1(\omega, T)$ are  correlated
with a relatively large in-plane superfluid density $n_s(0)$ at $T = 0$.

In this paper, we show that if the phases are assumed to fluctuate
{\em classically}, then the presence of {\em quenched disorder} will
produce a low-temperature background in $\sigma_1(\omega)$ which has
many similarities to that observed in experiment\cite{orenstein}.
Specifically, we argue that in a sample with  quenched disorder
there must inevitably be a low-temperature background $\sigma_1(\omega, T)$ 
arising from classical phase fluctuations, whose frequency integral is 
related to the zero-temperature superfluid density, $n_s(0)$.  While the
exact frequency dependence of this  background depends on the particular 
dynamics obeyed by the phase fluctuations, 
the frequency integral is independent of the dynamics, but depends
only on the assumption that the phases fluctuate {\em classically}. 
These fluctuations thus provide an alternative possible source of conductivity
background, in addition to the gapless quasiparticles which should exist
in a d-wave superconductor\cite{wen}. 

In order to model the phase fluctuations, we adopt a {\em classical XY model}
on a tetragonal lattice with different couplings in the $ab$ and $c$ 
directions\cite{carlson}. We consider a lattice model of a superconductor, 
such that each lattice point $i$ is characterized by a phase $\theta_i$.  
In the presence of a vector potential, the interaction Hamiltonian for
this system is given by 
\begin{eqnarray}
H = -\sum_{\langle ij;  \|\rangle}J_{ij;\|}\cos(\theta_i-\theta_j +A_{ij})
 \nonumber \\
-\sum_{\langle k\ell; \perp \rangle}J_{k\ell;\perp}\cos(\theta_k-\theta_\ell
+A_{k\ell}),
\label{eq:hamil}
\end{eqnarray}
where the first sum runs over nearest neighbors in the $ab$ plane and the
second over bonds in the $c$ direction; $J_{ij\|}$ and $J_{ij;\perp}$ are the
couplings between lattice points in the $ab$ and $c$ directions.
The gauge factor is
$A_{ij} = (2\pi/\Phi_0)\int_i^j{\bf A}\cdot{\bf d\ell}$, 
%CHANGE deleted ``with a similar expression for $A_{k\ell}$,''
where ${\bf A}$ is the vector potential, 
$\Phi_0 = hc/q$ is the flux quantum, and $q$ is the magnitude of the
charge of a Cooper pair.

The diagonal components $n_{s;\alpha\alpha}$ of the superfluid density
tensor are given by $n_{s;\alpha\alpha}  =
[(m^*c^2/(Vq^2)](\partial^2 F/\partial A_\alpha^2)_{A_\alpha = 0}$.
Here $m^*$ is the mass of a Cooper pair, $F$ the Helmholtz free energy,
$V$ the sample volume, $c$ the speed of light, and $A_\alpha$ is a fictitious
uniform vector potential applied in the $\alpha$ direction in the presence of
periodic boundary conditions in all three directions.
Using $F = -k_BT \ln \Pi_{i=1}^N \int_0^{2\pi}d\theta_i\exp(-H/k_BT)$, 
where $N$ is the number of lattice points, we obtain
\begin{equation}
n_{s;\alpha\alpha} = \frac{m^*c^2}{Vq^2}(\gamma_{1;\alpha} - \gamma_{2;\alpha}).
\label{eq:nsalpha}
\end{equation}
Here $\gamma_{1;\alpha} = [(2\pi/\Phi_0)a_\alpha]^2E_\alpha$, where
$a_\alpha$ is the lattice constant in direction $\alpha$ 
($a_x = a_y \neq a_z$ for the tetragonal lattice),
$\Phi_0=hc/q$ is the flux quantum, and
$E_\alpha =\langle \sum_{\langle ij\rangle \| \alpha}
J_{ij}\cos\theta_{ij}\rangle$, the sum running over
bonds in the direction $\alpha$.   Similarly, 
$\gamma_{2;\alpha} = [V^2/(c^2k_BT)]\langle {\cal J}_\alpha(t)^2\rangle$, where
${\cal J}_\alpha(t) = (a_\alpha/V)\sum_{\langle ij\rangle \| \alpha}
(qJ_{ij}/\hbar)\sin\theta_{ij}$,
$\theta_{ij} = \theta_i-\theta_j$, the sums are carried out over
distinct bonds in the $\alpha$ direction, and the triangular brackets
denote a canonical thermal average.

The key point is to note that ${\cal J}_\alpha(t)$ is the volume-averaged
supercurrent density in direction $\alpha$.   From this connection, we can
use the Kubo formula\cite{mahan}, in the classical limit, together with
eq.\ (\ref{eq:nsalpha}), to obtain an expression for 
$\sigma_{1;\alpha\alpha}(\omega)$, the real 
part of the long-wavelength conductivity in the direction $\alpha$: 
\begin{eqnarray}
\sigma_{1;\alpha\alpha}(\omega) &=& 
\frac{V}{k_BT}\int_0^\infty dt \cos(\omega t)
\langle {\cal J}_\alpha(t){\cal J}_\alpha(0)\rangle \nonumber\\
&=& \frac{a_\alpha^2}{Vk_BT}
\int_0^\infty dt\cos(\omega t) \langle {\cal I}_{\alpha}(0){\cal I}_\alpha
(t)\rangle,
\label{eq:sigalpha}
\end{eqnarray}
where $I_{c;ij} = qJ_{ij}/\hbar$ and 
\begin{equation}
{\cal I}_\alpha(t)= \sum_{\langle ij \rangle \| \alpha}
I_{c;ij}\sin\theta_{ij}(t).
\label{eq:ialpha}
\end{equation}
Integrating eq.\ (\ref{eq:sigalpha}) over frequency leads to
$\int_0^\infty\sigma_{1;\alpha\alpha}(\omega)d\omega
= [V\pi/(2k_BT)]\langle {\cal J}_{\alpha}^2\rangle$.
Substituting this expression into eq.\ (\ref{eq:nsalpha}), and using
$\sigma_{\alpha\alpha}(-\omega)=\sigma_{\alpha\alpha}(\omega)$,
we finally obtain
\begin{eqnarray}
S_{1;\alpha} 
&\equiv& \int_0^\infty\sigma_{1;\alpha\alpha}(\omega)d\omega \nonumber\\
&=&\frac{\pi c^2}{2}\left[
 \left(\frac{2\pi a_\alpha}{\Phi_0}\right)^2\frac{E_\alpha}{V}
 -\frac{n_{s;\alpha\alpha} q^2}{m^* c^2}
\right].
\label{eq:fluc}
\end{eqnarray}

We first show that, for an {\em ordered} system, 
${S}_{1;\alpha}^{o}(T)= 0$ at $T=0$.
Suppose that $J_{ij} = J_\alpha$ for all bonds in the $\alpha$ direction.
Then the phases $\theta_i$ are all equal and, with periodic boundary
conditions, $E_\alpha = NJ_\alpha$.  Similarly, in the limit $T \rightarrow 0$, 
$(\partial^2 F/\partial A_\alpha^2)_{A_\alpha = 0} =
(\partial^2 H/\partial A_\alpha^2)_{A_\alpha = 0} 
= (2\pi a_\alpha/\Phi_0)^2 NJ_\alpha$, 
from direct evaluation of the second derivative, with periodic
boundary conditions.  Hence, both terms on the right-hand side
of eq.\ (\ref{eq:fluc}) are equal, and $S_{1;\alpha}^{o}(0) = 0$.

We can use this result to obtain some analytical results for 
$S_{1;\alpha}(T = 0)$ in the presence of quenched disorder.  
Since the right-hand side of eq.\ (\ref{eq:fluc}) 
vanishes for the ordered case,
%CHANGE  ...(T=0)  ->  ...(0)
$n_{s:\alpha\alpha}^o(0)q^2/m^*c^2 = 
(2\pi a_\alpha/\Phi_0)^2E_{\alpha}^o(0)/V$,
where the superscript refers to the ordered system.  Using this
equivalence, we may rewrite $S_{1;\alpha}(T)$ as
\begin{equation}
%CHANGE
S_{1;\alpha}(T) =
\frac{\pi q^2n_{s;\alpha\alpha}^o(0)}{2m^*}
\left(\frac{E_\alpha(T)}{E_\alpha^o(0)}-\frac{n_{s;\alpha\alpha}(T)}
{n_{s;\alpha\alpha}^o(0)}\right).
\label{eq:s1}
\end{equation}
If the disordered system is unfrustrated (all $J_{ij} >0$), 
all $\theta_i$ are still equal at $T=0$, and $E_\alpha(T=0)$ is
controlled by the average bond strength $\bar{J}_\alpha$: 
%CHANGE
$E_\alpha(0) = N\bar{J}_\alpha$.  In view of this fact, we choose  
reference system to have $J_\alpha^o=\bar{J}_\alpha$, so that
$E_\alpha(0)/E_\alpha^o(0) = 1$.

To further analyze this regime, we use a generalization of
a theorem proved by Kirkpatrick \cite{kirkpat}, which maps the spin-wave
stiffness constant of a random Heisenberg ferromagnet onto the conductance
of a disordered conductance network, and generalized
to the superfluid density of $XY$ systems (the analog of the
spin-wave stiffness constant) in Ref.\ \cite{ebner83}.
The generalized theorem, as applied to the present anisotropic geometry,
states that
\begin{equation}
\frac{n_{s;\alpha\alpha}(T = 0)}{n_{s;\alpha\alpha}^o(T=0)} 
= \frac{g_{\alpha\alpha}}{g_{\alpha\alpha}^o}.
\label{eq:kirkp}
\end{equation}
Here $g_{\alpha\alpha}$ and $g_{\alpha\alpha}^o$ are components of the 
conductivity tensor of 
fictitious random conductance networks in which
the bond conductances are $J_{ij}$ and $J_{ij}^o$, the bond strengths of
the real and reference systems.

Using this theorem, we can evaluate $S_{1;\alpha}(0)$ in an isotropic
system for weak disorder.  We retain our choice of the $\bar{J}_\alpha$ for
the reference system.  For an isotropic conductance network in $d$ 
dimensions\cite{bergman}, 
$g/\bar{g} = 1 - \langle (\delta g)^2\rangle_{dis}/[d\bar{g}^2] +{\cal O}(
(\delta g)^3)$, where $\delta g = g - \bar{g}$ and $\langle...\rangle_{dis}$
denotes an average over configurations of the quenched disorder.
Hence, making use of eq.\ (\ref{eq:kirkp}), we obtain
$n_{s;\alpha\alpha}(0) = n_{s;\alpha\alpha}^o(0)
\left(1 -\frac{\langle (J - \bar{J})^2\rangle_{dis}}{\bar{J}^2 d}\right)$,
where $d = 2$ or $3$ is the dimensionality and $\bar{J}$ is the average
bond strength of a bond in this network.

Substituting this expression back into eq.\ (\ref{eq:s1}) gives 
our final result for the integrated fluctuation conductivity in the case of
weak disorder:
\begin{equation}
%CHANGE (t=0)
\int_0^\infty\sigma_{1;\alpha\alpha}(\omega, T=0)d\omega
=\frac{\pi q^2n_{s;\alpha\alpha}(0)}{2m^*}\frac{\langle (J -
\bar{J})^2\rangle_{dis}}{\bar{J}^2d}.
\label{eq:fluc1}
\end{equation}
Here we have used the fact that, $n_{s;\alpha\alpha}(0)$ and
$n_{s;\alpha\alpha}^o(0)$ are equal to lowest order in 
$\langle(J - \bar{J})^2\rangle_{dis}$.
The dimensions $d = 2$ or $d = 3$ 
would be appropriate for the limiting cases of no interplanar
coupling (d = 2) or a fully isotropic system (d = 3); presumably a material
such as \bscco would behave similarly to the $d = 2$ case.

The result (\ref{eq:fluc1}) has several similarities to 
experiments\cite{corson,orenstein}.  Most strikingly, at fixed magnitude of the
disorder parameter $\langle (J - \bar{J})^2\rangle_{dis}/\bar{J}^2$,
$S_{1;\alpha}(T = 0)$ is predicted to scale with the low-temperature superfluid
density, in agreement with experiment.  Of course, the experiment is carried
out at a specific frequency, while the relationship (\ref{eq:fluc1})
applies to a frequency {\em integral}, but presumably
$\sigma_{1;\alpha\alpha}(\omega)$ behave similarly for any given
frequency.  Note also that the result does not require the presence or
absence of short range order in the bond strength\cite{bergman}, but only
that the bond distribution of bonds be macroscopically isotropic
in $d$ dimensions.

It is of interest to make a numerical estimate of the integral 
(\ref{eq:fluc1}) for parameters appropriate for \bscco.  From the London
equations, we may write $\pi q^2n_{s;\alpha\alpha}/(2m^*) = c^2/(8\lambda^2)$.
If we take the in-plane penetration depth as $2000 \AA$ and we use a
low-temperature conductivity $\sigma_1(\omega, 0)$ of about $10^6$ 
$\Omega^{-1}$m$^{-1}$ at $\omega \sim 140$ GHz,
as suggested in Ref.\ \cite{orenstein}, and we further assume that this
value remains roughly constant to low frequencies and cuts off at only
slightly higher frequencies, then relation (\ref{eq:fluc1}) could be satisfied
for $\langle (J - \bar{J})^2\rangle/\bar{J}^2 \sim 10^{-1}$.  This is at best
an order-of-magnitude estimate, but it suggests at least that the relation
(\ref{eq:fluc1}) is not ruled out by experiment, since mean-square fluctuations
in this range seem reasonable.

The present results are entirely independent
of the specific dynamics: {\em any} classical dynamics will give the same 
value for $S_{1;\alpha}(T)$.  Nevertheless, in order to illustrate
a possible behavior for $\sigma_1(\omega)$, we have approximated the
dynamical response of the phase by the equations describing a 
coupled array of overdamped Josephson junctions [see, for example,
Ref. (\cite{hwang})].  In this picture, each lattice
point is connected by an overdamped Josephson junction which carries three
current contributions in parallel: a supercurrent $I_{c;ij}\sin\theta_{ij}$,
a normal current through a shunt resistance $R_{ij}$, and a Langevin
noise current $I_{L;ij}(t)$ representing the effects of thermal fluctuations.

We will consider the anisotropic limit, in which the bond
strengths $J_{ij}$ vanish for $ij \| c$. In this 
limit, the supercurrents from bonds in different layers are uncorrelated and
eq.\ (\ref{eq:sigalpha}) reduces to (for conductivity 
$\sigma_{1;xx}(\omega, T)$ parallel to the $a$ axis)
\begin{equation}
\sigma_{1;xx}(\omega) = \frac{1}{N_sa_zk_BT}\int_0^\infty dt \cos(\omega t)
\langle {\cal I}_\alpha(0){\cal I}_\alpha(t)\rangle,
\label{eq:cond}
\end{equation}
where $N_s$ is the number of lattice points in one layer, and the sum defining
${\cal I}_\alpha$ [eq.\  (\ref{eq:ialpha})] runs over
lattice points in a {\em single} layer.  
We have evaluated this average by solving the coupled
Josephson equations, using the method described in ref.\ \cite{hwang}.  

Figs.\ 1 and 2 show $\sigma_{1;xx}(\omega, T)$ for this model in two cases.  
In Fig.\ 1, 
all the critical currents and shunt resistance have the same values, $I_c$ 
and $R$.  In Fig.\ 2, 
the $I_{c;ij}$ are uniformly and independently distributed on the 
interval $(0, 2I_c)$, but all the $R$'s remain identical.
In both cases, time is expressed in units of
$\hbar/(qRI_c)$, frequency in units of $qRI_c/\hbar$, current in units of 
$I_c$, $k_BT$ in units of $\hbar I_c/q$, and therefore
$\sigma_{1;xx}(\omega, T)$ in units of $1/(Ra_z)$.

The results for the two cases are strikingly different.  For the {\it ordered}
lattice, there is a fluctuation peak in $\sigma_1(\omega, T)$, 
more prominent at smaller frequencies, centered near
the lattice phase-ordering temperature $T_c \sim 0.95 \hbar I_c/q$\cite{tc}.
[The peak is probably shifted away from this value by finite-size effects
in our calculations.]  For $T <T_c$, at all values of $\omega$,
$\sigma_1(\omega, T)$ falls off sharply towards a very small value at $T = 0$,
consistent with the prediction that $S_{1;\alpha\alpha}(0) = 0$.
[The decreasing character of $\sigma_1(\omega, T)$ is most evident 
in the semilog plot.]

\begin{figure}[t]
\vspace{0.5in}
\leftline{ \psfig{file=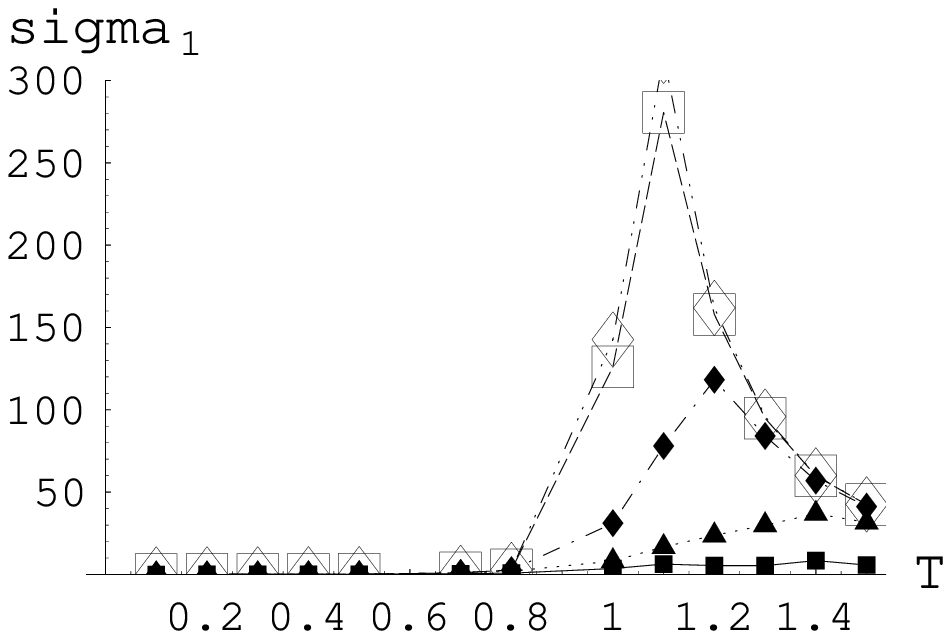,width=3.4in} }
\leftline{ \vbox{ \psfig{file=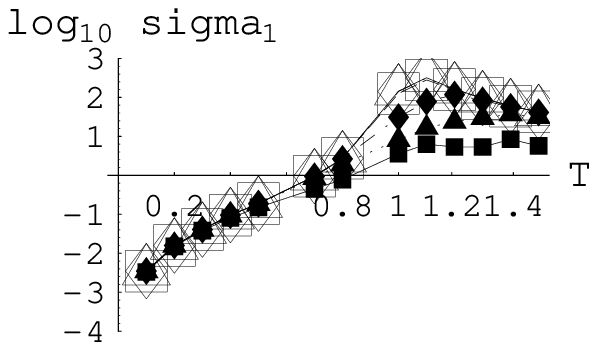,width=2.35in}
 \psfig{file=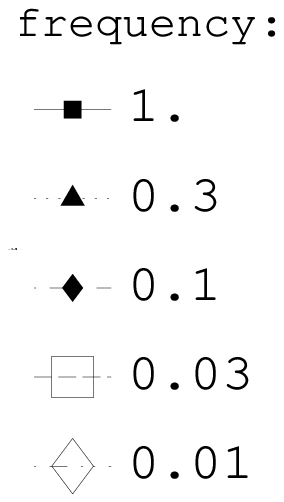,width=.8in} 
}}

\label{fig1}

\vspace{0.2in}
\caption[]
{
%CHANGE
Fluctuation conductivity $\sigma_1(\omega, T)$, plotted as a function of
temperature $T$ for several values of the frequency $\omega$, calculated
for an {\it ordered} $32 \times 32$ lattice with periodic boundary conditions.
The inset shows same results in a semilog scale, demonstrating that
$\sigma_1(\omega, T \rightarrow 0) \rightarrow 0$.
Units are as described in the text, and lines are guides to the eye.}
\end{figure}

In the {\it disordered} lattice, $\sigma_1(\omega, T)$ 
still has a strong fluctuation peak, but in addition has a broad
background for all $T < T_c$, which rolls off at
larger frequencies.  The frequency-dependence is probably Lorentzian, as would
be expected if the current-current correlation function in 
(\ref{eq:sigalpha}) decays exponentially in time.
We expect that the relaxation time $\tau(T)$ should
be of order $\hbar/(2eRI_c)$.  The semilog plot shows clearly that
 $\sigma_1(\omega, T)$ remains finite even as $T \rightarrow 0$. 

The disorder-induced broad background in our calculations may appear
rather small (much weaker than the peak near $T_c$).
But this background is calculated in units of the strongly
temperature-dependent coupling energy $J$.  According
to one model, for example\cite{ebner83}, $J \propto \Delta^2$, where
$\Delta$ is the mean-field energy gap, a quantity which decreases to
zero at the mean-field transition temperature $T_{c0}$.  When the
temperature-dependence of $J$ is properly included, the fluctuation peak may
well prove to be much smaller than the low-temperature background, especially
when the low-temperature $n_s$ is relatively large; this behavior would
be in agreement with experiment\cite{orenstein}.      

\noindent
\begin{figure}[ht]
\vspace{0.5in}
\leftline{ \psfig{file=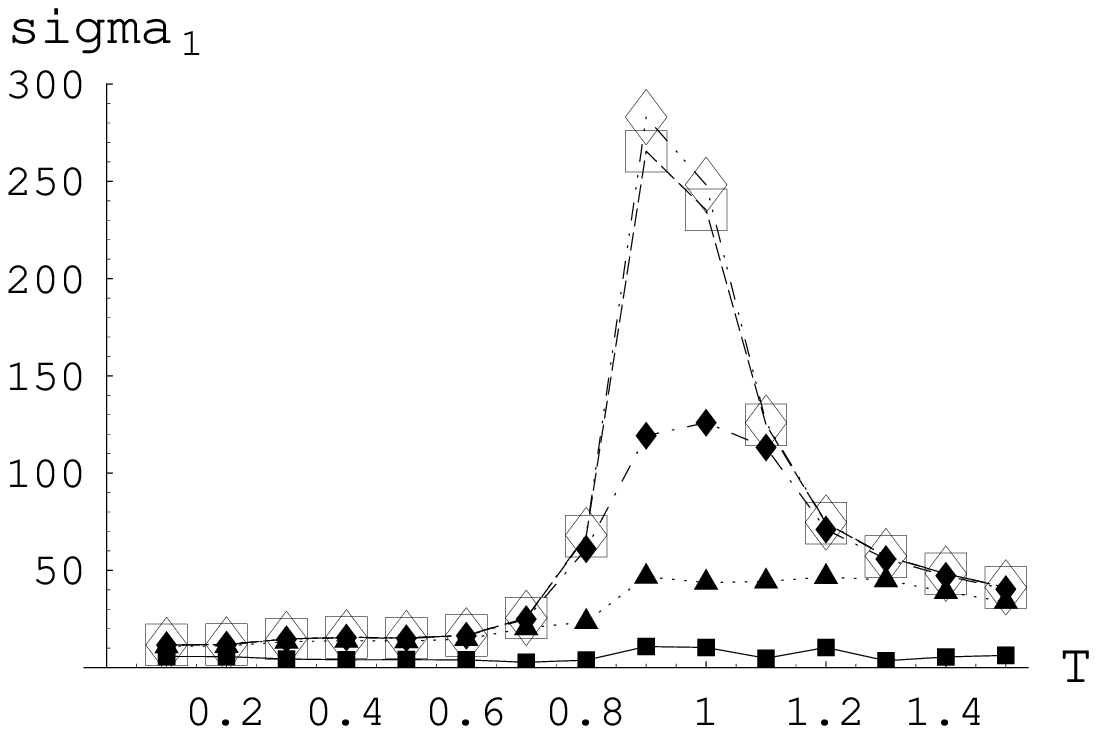,width=3.2in} }
\leftline{ \vbox{ \psfig{file=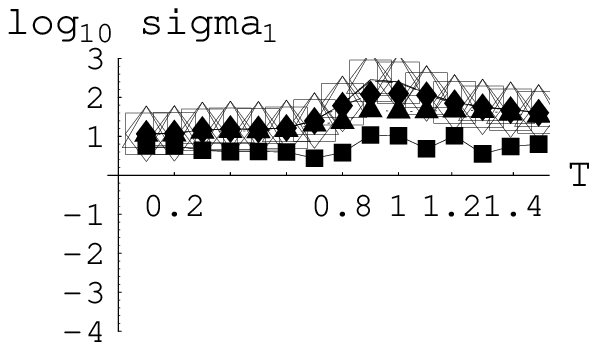,width=2.35in}
 \psfig{file=Legend.eps,width=.8in} 
}}
\label{fig2}

\vspace{0.2in}
\caption[]
{
%CHANGE
$\sigma_1(\omega, T)$ plotted as a function of $T$ for several frequencies
in a {\it disordered} lattice ($32 \times 32$ array), 
in which the critical current
is uniformly and randomly distributed on the interval $(0, 2I_c)$. 
The inset shows same results in a semilog scale.
Lines are guides for eyes.
}
\end{figure}
\vspace{0.1in}

We believe that the disorder-induced background conductivity
originates as follows.  In the absence of disorder, 
phase fluctuations at low $T$ will occur in the form of long-wavelength,
low-frequency ``phase phonons.''   These are loosely analogous to spin waves  
in ferromagnets, but may be underdamped or
overdamped, depending on the dynamics.   In the presence of quenched disorder,
the density of states of these phase phonons is affected
by the scattering of these excitations off the disorder; in addition, the
modes of a given frequency, instead of having a unique wave vector, will
form a wave packet with a spread in wave vector.
Analogous behavior is found in the spin wave spectra of ferromagnets
when there is quenched disorder in, e. g., the exchange constants.
We speculate that these damped modes produce the
extra contribution to the $\sigma_1(\omega, T)$ at low temperatures. 

There are several issues which we have not considered in the present
work.  For example, classical phase fluctuations might be expected to
be frozen out by quantum effects at low $T$, as may happen in low-$T_c$
superconductors\cite{lemberger}.  But in some high-T$_c$ materials,
the existence of nodal quasiparticles may provide a normal background which
would impede this quantum freeze-out.  It remains an open question
to what temperature classical phase fluctuations persist in such materials
as \bscco.   

In conclusion, we have demonstrated, both analytically and numerically,
that in a superconductor with  {\it quenched disorder}, 
the finite-frequency electrical conductivity $\sigma_1(\omega, T)$ 
remains non-zero at low temperatures if the order parameter has phase
fluctuations which can be treated classically. 
If this assumption is satisfied, this result is quite general,
and independent of the dynamical equations
obeyed by these fluctuation.  For weak disorder,
the frequency integral of this fluctuation conductivity scales 
proportional to the low-temperature superfluid density, in agreement
with recent microwave experiments in \bscco.

\section{Acknowledgments}

This work was supported by the National Science Foundation, Grant
DMR97-31511.  We are grateful for valuable conversations with J. Orenstein,
D.\ J.\ Bergman and T.\ R.\ Lemberger.

%\newpage

% \end{multicols}

%\normalsize
%\vspace{0.2in}

%\noindent
%the figure was here
\noindent
\vspace{0.1in}  

%\end{twocolumn}


\begin{thebibliography}{3}

\bibitem{nelson} D. R. Nelson, Phys. Rev. Lett. {\bf 60}, 1973 (1988).

\bibitem{fisher} D. S. Fisher, M. P. A. Fisher, and D. A. Huse,
Phys. Rev. {\bf B43}, 130 (1991).

\bibitem{salamon} M. B. Salamon, J. Shih, N. Overend, and M. A. Howson,
Phys. Rev. {\bf B47}, 5523 (1993).

\bibitem{emery} V. J. Emery and S. A. Kivelson, Nature {\bf 374}, 434 (1995).

\bibitem{anlage} S. M. Anlage, J. Mao, J. C. Booth, D. H. Wu, and
J. L. Peng, Phys. Rev. {\bf B53}, 2792 (1996).

\bibitem{corson} J.\ Corson {\it et al}, Nature {\bf 398}, 221 (1999).

\bibitem{lowfreq} S.-F.\ Lee {\it et al}, Phys. Rev. Lett. {\bf 77}, 
735 (1996).

\bibitem{orenstein} J.\ Corson, J.\ Orenstein, J.\ N.\ Eckstein, and
I.\ Bozovic, cond-mat/9908368.

\bibitem{wen} See, for example, X. G. Wen and P. A. Lee, Phys. Rev. Lett.
{\bf 80}, 2193 (1998); M. Franz and A. J. Millis, Phys. Rev. {\bf B58}, 14572
(1998). 

\bibitem{carlson} E. W. Carlson, S. A. Kivelson, V. J. Emery, and E.
Manousakis, Phys. Rev. Lett. {\bf 83}, 612 (1999).

\bibitem{mahan} See, e.\ g.\, G.\ Mahan, {\em Many-Particle Physics}
(Plenum, New York, 1981), p. 194.

%\bibitem{roddick} E. Roddick and D. Stroud, Phys. Rev. Lett. {\bf 74},
%1430 (1995).

\bibitem{kirkpat} S.\ Kirkpatrick, Rev.\ Mod.\ Phys. {\bf 45}, 573 (1973).

\bibitem{ebner83} C. Ebner and D. Stroud, Phys. Rev. {\bf B28}, 5053 (1983).

\bibitem{bergman} See, for example, D. J. Bergman and D. Stroud, Solid
State Physics {\bf 46}, 147 (1992).

\bibitem{hwang} I.\ J.\ Hwang and D. Stroud, Phys. Rev. {\bf B57}, 6036 (1998).

\bibitem{tc} P. Olsson, Phys. Rev. {\bf B52}, 4511 (1995).

\bibitem{lemberger} S.\ J.\ Turneaure, T.\ R.\ Lemberger, and J.\ M.\
Graybeal, Phys.\ Rev.\ Lett. {\bf 84}, 987 (2000).

%\bibitem{paget} K.\ M.\ Paget, B.\ R.\ Boyce, and T.\ R.\ Lemberger, 
%Phys. Rev. {\bf B59}, 6545 (1999).


\end{thebibliography}
\end{document}